\documentclass[a4paper,12pt]{article}
\usepackage{natbib}
\usepackage[english]{babel}
\usepackage{setspace}
\usepackage[utf8x]{inputenc}
\usepackage{graphicx} 
\usepackage[left=2cm,right=2cm,top=2cm,bottom=2cm]{geometry}
\usepackage{textcomp}
\usepackage[T1]{fontenc}
\usepackage{amssymb}
\usepackage{xcolor}
\usepackage{authblk}
\usepackage{ulem}
\usepackage{hyperref}
\usepackage{mathabx}
\usepackage{appendix}

\title{
\large
\textbf{3D convection-resolving model of temperate, tidally-locked exoplanets}
}

\author[1]{Maxence Lef{\`e}vre}
\author[2]{Martin Turbet}
\author[1]{Raymond Pierrehumbert}
\affil[1]{Department of Physics (Atmospheric, Oceanic and Planetary Physics), University of Oxford, Parks Rd, Oxford, OX1 3PU, UK}
\affil[2]{Observatoire Astronomique de l'Universit{\'e} de Gen{\`e}ve, 51 chemin de P{\'e}gase, 1290 Sauverny, Switzerland}

\date{}

\begin{document}

\maketitle
\newpage
\section*{Abstract}

A large fraction of known terrestrial-size exoplanets located in the Habitable Zone of M-dwarfs are expected to be tidally-locked. Numerous efforts have been conducted to study the climate of such planets, using in particular 3-D Global Climate Models (GCM). One of the biggest challenges in simulating such an extreme environment is to properly represent the effects of sub-grid convection. Most GCMs use either a simplistic convective-adjustment parametrization or sophisticated (e.g., mass flux scheme) Earth-tuned parametrizations. One way to improve the representation of convection is to study convection using Convection Resolving numerical Models (CRMs), with an fine spatial resolution . In this study, we developed a CRM coupling the non-hydrostatic dynamical core WRF with the radiative transfer and cloud/precipitation models of the LMD-Generic climate model to study convection and clouds on tidally-locked planets, with a focus on Proxima b. Simulations were performed for a set of 3 surface temperatures (corresponding to three different incident fluxes) and 2 rotation rates, assuming an Earth-like atmosphere. The main result of our study is that while we recover the prediction of GCMs that (low-altitude) cloud albedo increases with increasing stellar flux, the cloud feedback is much weaker due to transient aggregation of convection leading to low partial cloud cover.

\section{Introduction}

The habitable zone (HZ) is classically defined as the range of orbital distances for which a planet can sustain water in liquid phase at the surface \cite{Kast93}, providing a potentially suitable environment (surface, liquid water, photons) for the emergence of life as we know it on Earth. Depending on the rotation state and the atmospheric composition of the planet, various processes can either extend or narrow the HZ. While the presence of greenhouse gases tends to increase the surface temperature, the radiative effect of clouds can be much more subtle. Clouds can scatter (and thus reflect) as well as absorb a significant part of incident stellar radiation. But they can also absorb (and scatter) the infrared thermal emission from the surface and the atmosphere, and then re-emit (or back-scatter, respectively) infrared thermal radiation back to the surface.

On Earth, the relative ratio between cloud-free and cloudy area in the tropics have a strong effect on the mean temperature \citep{Pier95}. It is therefore crucial to know the composition, thickness, vertical and horizontal spatial distributions of these clouds to correctly evaluate the climate state of the Earth and by extension of other planets in or outside of the solar system. 

Most terrestrial-sized exoplanets located in the HZ that have been detected so far \citep{Angl16,Gillo17,Bonf18,Zech19,Tuom19} are orbiting around low mass stars, also known as M-stars, the most abundant and longest-lived stars in the Milky Way \citep{Kirk12}. A significant number of the planets orbiting in the Habitable Zone of low mass stars are expected to have near-zero obliquity and be in synchronous rotation around their host star (i.e. with one side of the planet permanently facing its host star). On a synchronously rotating telluric exoplanet covered with water, the constant stellar radiation received at the substellar point should create a strong convective region producing very thick clouds reflecting the stellar light \citep{Yang13}. This behaviour has been shown to be robust across a wide range of 3-dimensional Global Climate Models \citep{Yang19}. The more irradiation a synchronous planet receives, the stronger the moist convection and thus the more reflective the substellar cloud cover is \citep{Yang13}. As a result, the inner edge of the HZ for synchronously rotating planets was found to lie significantly closer to the host star \citep{Yang13}. This mechanism may be less effective for synchronous planets in the fast-rotating regime, i.e. with orbital periods roughly lower than 10~days \citep{Edso11,Haqq-Misra18}. For these planets, winds are strong enough to shift the clouds away from the substellar point \citep{Kopp16}, thus reducing the effective cloud coverage in the substellar region, where the peak of instellation is. This effect weakens the cloud feedback described above \citep{Kopp16}. This effect highlights the importance of knowing accurately the composition, thickness, as well as the vertical and horizontal distributions of clouds on a synchronously rotating planet to get insight on its mean climate state.

The representation of clouds and turbulence in the Global Climate Models (GCMs) is one of the most important uncertainties of these models. In fact, the typical size of the structures of these phenomena, i.e. the shallow convection cells diameter, varies on Earth from 10 to 40~km \citep{Atki96}. This is much lower than the typical horizontal resolution of GCMs ($\sim$~100~km). GCMs use therefore sub-grid parameterizations to represent the effects of convection and turbulence, which is a main source of uncertainty in the models \citep{Yang19,Fauchez:2019gmd}. In fact, these schemes are either (i) too simple to represent the mixing of wind, energy and tracer by the convection, or either (ii) sophisticated, but tuned to simulate present-day Earth, and can therefore give inaccurate results when used to simulate environments very different from the Earth, such as planets in synchronous rotation. To be able to resolve convection in such exotic environments (in order to better understand them), it is necessary to run models at a spatial resolution similar to that at which convection processes operate.

Correctly modeling convection on synchronously rotating planets is not only crucial to their mean climate state and thus their potential habitability, but is also key to the observability of these planets in reflected light. Direct imaging in reflected light is indeed one of the most promising avenue to detect and characterize temperate, Earth-size planets orbiting around nearby stars. Specifically, it has been shown that the reflected light of very nearby temperate planets such as Proxima~b \citep{Angl16} could be detected and analysed with ELT-class telescopes \citep{Turb16}. \citet{Lovi17} even proposed that, combining direct imaging with high-resolution spectroscopy on a 8~m-class telescope, observations of Proxima~b, in particular in reflected light could be attempted. However, our ability to detect and characterize Proxima~b (and its atmosphere, surface, and clouds) depends strongly on the amount of light reflected by the planet. This amount highly depends on the reflectivity of the cloud cover (especially near and eastward of the substellar region), where the signal measured by direct imaging is more favorable.

On Earth intense convective activity occurs in the tropics leading to a cloud coverage of about 70$\%$ above the oceans \citep{Eas11,Stub06}. Several types of clouds play a distinct role in that net coverage. In the tropics the low altitude stratocumulus covers 20~$\%$ of the surface, mesoscale high opaque clouds 6~$\%$ and cirrus clouds about 45~$\%$. Since the 80's, some models were developed with the aim to study the convection \citep{Lipp86,Lipp88} and its organization \citep{Held93}. With evolution of computing capacity, the model evolved from two-dimensional models \citep{Grab01,Grab02}, to three-dimensional channels configuration models \citep{Tomp01,Wing16}, and last to periodic three-dimensional models \citep{Tomp01,Mull12,Wing14}. From observations campaign \citep{Tobi13} and numerical modeling, several convective and squall line and self-aggregation clusters where clouds occupy only a limited area leading to a drier free troposphere and larger domain-mean outgoing longwave radiation \citep{Wing17}. Self-aggregation is also obtained in aquaplanet with GCMs and parameterized convection \citep{Bony16}, resulting to a control of the anvil cloud fraction by the sea surface temperature (SST), a warmer SST leads to less anvil cloud fraction.

\cite{Zhan17} performed the first modelling effort to study the convective activity in synchronously rotating planets orbiting around M-stars. With a 3~km resolution and GCM generated boundary conditions, using radiative transfer and microphysics designed for Earth study this model investigated only one incident flux value of 2000 W~m$^{-2}$. \cite{Serg20} studied as well the convection on Proxima Centauri b using a GCM with a zoom at the substellar point, lowering the resolution down to 4~km covering 40 degrees of latitude and longitude. \cite{Serg20} focused on the impact of resolved convection on the large scale dynamics for one single stellar flux and rotation rate and also on the comparison between the resolved convection and two different GCM convection parameterizations. Here we propose to study the convection regime of tidally-locked environment using convection-resolving modelling with a realistic radiative transfer and water cloud microphysics. Most simulations were run for the case study of the temperate, terrestrial-mass exoplanet Proxima~b, because we were motivated by its high potential for future direct imaging observations. However, we explore broader cases than just Proxima~b, by first varying the spectral type of the host star, then varying the strength of the stellar flux and last varying the rotation period of the planet (to recover cases simulated in \citealt{Yang13}). The main goal of our study is to understand how the convection behaves at the substellar point of aquaplanets in synchronous rotation. This is a promising pathway to derive realistic parameterizations of the convection as well as the clouds albedo impacting bond albedo of such planets.

The paper is organized as follows. Our CRM model is described in Section~\ref{Sec:model}. In Section~\ref{Sec:ref}, the reference simulation is presented. The impact of the incident stellar flux and rotation rate are discussed in Sections \ref{Sec:flu} and \ref{Sec:rot}, respectively. The results are discussed in Section~\ref{Disc} and our conclusions are summarized in Section~\ref{Conc}.

\section{The LMD Generic CRM model}
\label{Sec:model}

\subsection{Dynamical core}

This study is conducted using the fully-compressible non-hydrostatic dynamical core of the Advanced Research Weather-Weather Research and Forecast (hereafter referred to as WRF) terrestrial model \citep{Skam08,Moen07}. The Large-Eddy Simulations (LES) mode is used: the grid spacing of the WRF model is refined to resolve the largest turbulent eddies responsible for most of the energy transport by buoyant convection, \citep{Lill62,Sull11}. Such modelling technique has been used to study Earth convection and small-scale turbulence. The atmospheric turbulence modelling has been conducted on various planets with WRF: on Mars \citep{Spig10}, on Venus \citep{Lefe17,Lefe18} and exoplanetary environment \citep{Zhan17}. Subgrid-scale ``prognostic Turbulent Kinetic Energy'' closure by \citet{Dear72} is used to parametrize turbulent mixing by unresolved small-scale eddies as in \citet{Moen07}.

\subsection{Coupling with complete LMD Generic GCM physical packages.}

The WRF core was coupled to the LMD Generic Model physics package in a similar way that the LMD Mars mesoscale model \citep{Spig09} and the LMD Venus Mesoscale model \citep{Lefe18,Lefe20}. Due to timescale of the convection and clouds dynamics, the LMD Generic CRM uses an online raditive transfer and is therefore a category 3 LES \citep[according to the terminology described in section 2.4 of][]{Spig16}. The LMD Generic physics is a versatile package that was used on various studies with the LMDz 3-D dynamical core from low irradiated planets terrestrial planets such as Archean Earth \citep{Char13} , Early Mars \citep{Forg13b,Word13,Turb17,Turbet:2019natsr,Turbet:2020impact}, Snowball Earth-like planets or exoplanets \citep{Turb17b} ; as well as for terrestrial exoplanets receiving a similar flux than Earth \citep{Word11,Bolm16,Turb16,Turb18,Aucl19} ; and for terrestrial exoplanets receiving relatively high irradiation such as future Earth \citep{Leco13b} or tidally locked exoplanets \citep{Leco13}. The LMD Generic physics was also used to simulate the atmosphere of solar system giant planets \citep{Spig20} and warm-Neptune-like exoplanets \citep{Char15b, Char15c}.

The radiative transfer of the LMD Generic physics package uses the correlated-k method \citep{Eyme16} for various species like CO$_2$, N$_2$ and H$_2$O. The radiative effect of clouds and Rayleigh scattering are taken into account. The incident stellar spectrum can be chosen to represent any type of host star. The spectrum used for Proxima Centauri (the star) was computed using the synthetic BT-Settl spectrum \citep{Rajp13} for a M~5.5 star, with T$_{eff}$~=~3000~K, g~=~10$^3$~m~s$^{-2}$ and [M/H]~=~0. 

Melting, freezing, condensation, evaporation, sublimation, and precipitation of H$_2$O are included in the model. Water cloud particle sizes are determined from the amount of condensed material and the number density of cloud condensation nuclei (CCN) set to 10$^6$~kg$^{-1}$ for liquid water clouds and to 10$^4$~kg$^{-1}$ for water ice clouds \citep{Leco13b}. Due to the small size of the grids in the CRM, typically from 10s~m to about 1~km, the cloud fraction of a grid cell can be only be either 0 or 100~\%. The sedimentation of ice particles and liquid droplets is computed following a modified Stokes law \citep{Ross78}. Precipitation of water is performed with the \cite{Bouc95} scheme. Evaporation of precipitation is also taken into account.

The purpose of this is paper to study the convection and clouds on temperate, tidally-locked aquaplanets with a focus on the exoplanet Proxima~b. From the radial velocity measurements of \citet{Angl16}, Proxima~b is likely a rocky planet with a most probable mass of 1.4 M$_\Earth$. We assumed that the density of the planet is similar to Earth (5500 kg~m$^{-3}$) and that the radius of the planet is equal to 7160~km (1.1~R$_\Earth$) as in \citet{Turb16}. The surface gravity is thus equal to 10.9~m~s$^{-2}$ for the fast rotation regime and 13.72~m~s$^{-2}$ for the slow rotation regime defined in the next subsection. The first value is the actual value of Proxima~b from the mass and radius measurements, and the second value is to facilitate the comparison with \cite{Yang14}. No obliquity and a circular orbit and a flat topography were assumed. The planet is assumed to be in synchronous rotation. The atmospheric composition is assumed to be close to the present-day Earth, i.e. N$_2$-dominated with 376 ppm of CO$_2$. The mixing ratio of H$_2$O can vary in space and time.

To ensure that the model is able to resolve realistic convective activity and clouds, it has first been tested using data from tropical convection collected during the Tropical Ocean-Global Atmosphere Coupled Ocean-Atmosphere Response Experiment \citep{Webs92}, aka the TOGA-COARE campaign. The convection resulting from the model (see Appendix~\ref{App}) consists of a shallow convection from the surface to 700~hPa (about 2.5~km) and deep convective plumes reaching 100~hPa (about 17~km). These altitudes are close to what has been measured during the campaign, and the associated surface rain over the simulated domain can reach up to 25 mm/day, close to the mean surface rain value observed in that TOGA COARE region \citep{John02} and in the tropics in general \citep{Kazu07}. With a simple microphysical model and configuration that are not tuned for Earth, the model is able to resolve realistic tropics convection. 

\subsection{Simulation setup}

\begin{center}
 \begin{figure}
 \includegraphics[width=17cm]{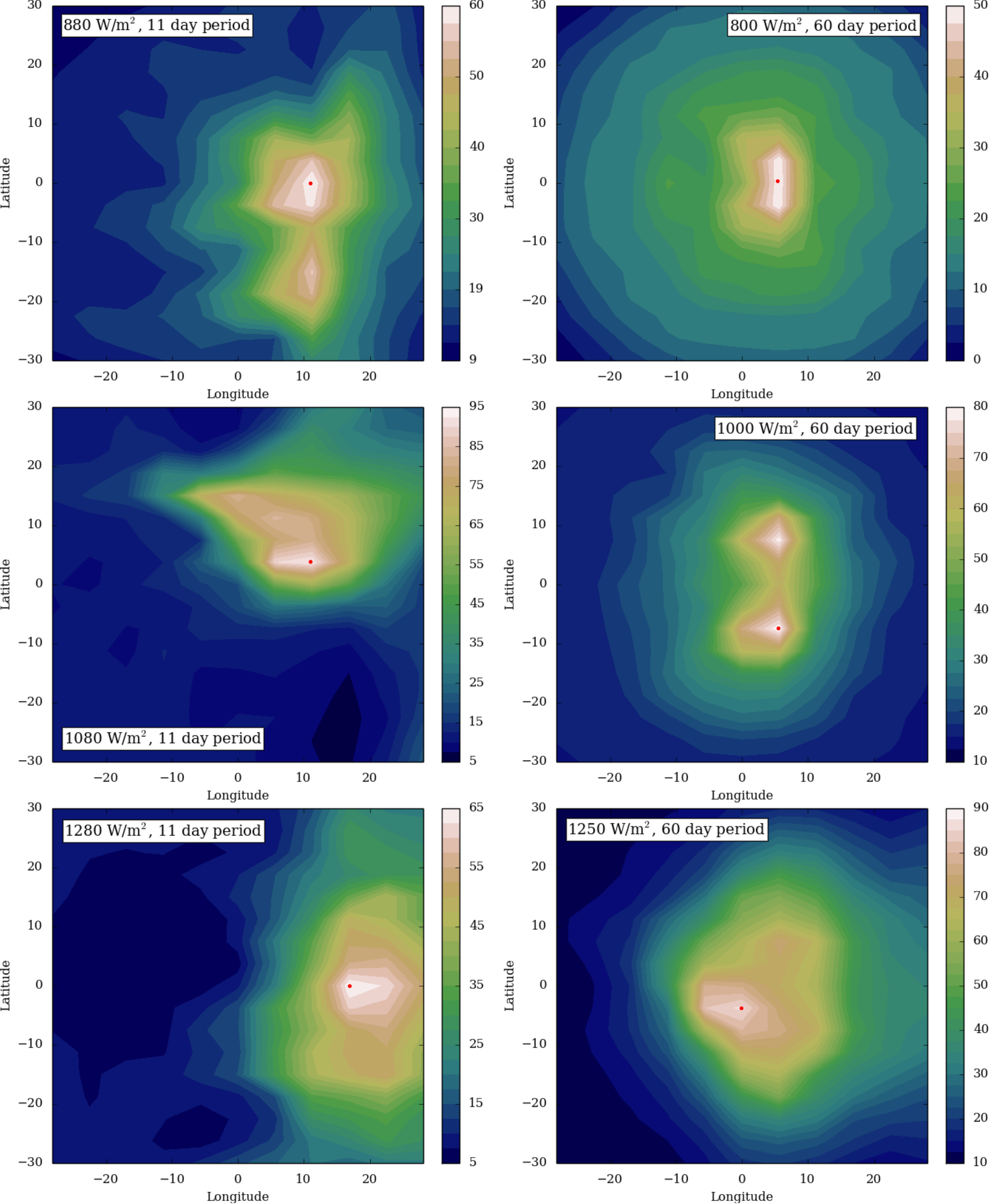}
 \caption{Vertically integrated time averaged over 20 days aerosol opacity in the GCM for an incident of 880W~m$^{-2}$ and 11 days rotation period (top left), 1080W~m$^{-2}$ and 11 days rotation period (middle left), 1280W~m$^{-2}$ and 11 days rotation period (bottom left), 800~W~m$^{-2}$ and 60 days rotation period (top right), 1080W~m$^{-2}$ and 60 days rotation period (middle right) and 1250W~m$^{-2}$ and 60 days rotation period (bottom right). The red dot  refers to the position where the initial profiles are extracted.}
 \label{21}
\end{figure}
\end{center}

The initial mean vertical profiles of atmospheric fields (temperature, pressure, winds and water) in the CRM were taken from LMD Generic GCM simulations of Proxima~b from \citet{Turb16}. Specifically, we extracted the mean vertical profiles at the substellar point, averaged over 20 consecutive Earth days. To explore different convective regimes, various incident stellar flux were considered, from 800 W~m to 1280~W~m$^{-2}$, as well as 2 rotation speeds, 6.3~10$^{-6}$ and 1.1~10$^{-6}$~s$^{-1}$ corresponding to rotation periods of 11 and 60~Earth days, respectively. The 11~days rotation period (relatively fast rotation regime) was chosen to fit the case of Proxima~b. The 60~days rotation period (slow rotation regime) was chosen to fit the main case of \citet{Yang13}. Note that in the GCM the maximum of the cloud fraction does not occur at the substellar point because of atmospheric dynamics and that this specific position may be different for another GCM. Fig~\ref{21} shows the vertically integrated time averaged cloud fraction in the GCM for the different cases of incident flux and rotation rate.

\begin{center}
 \begin{figure}
 \includegraphics[width=17cm]{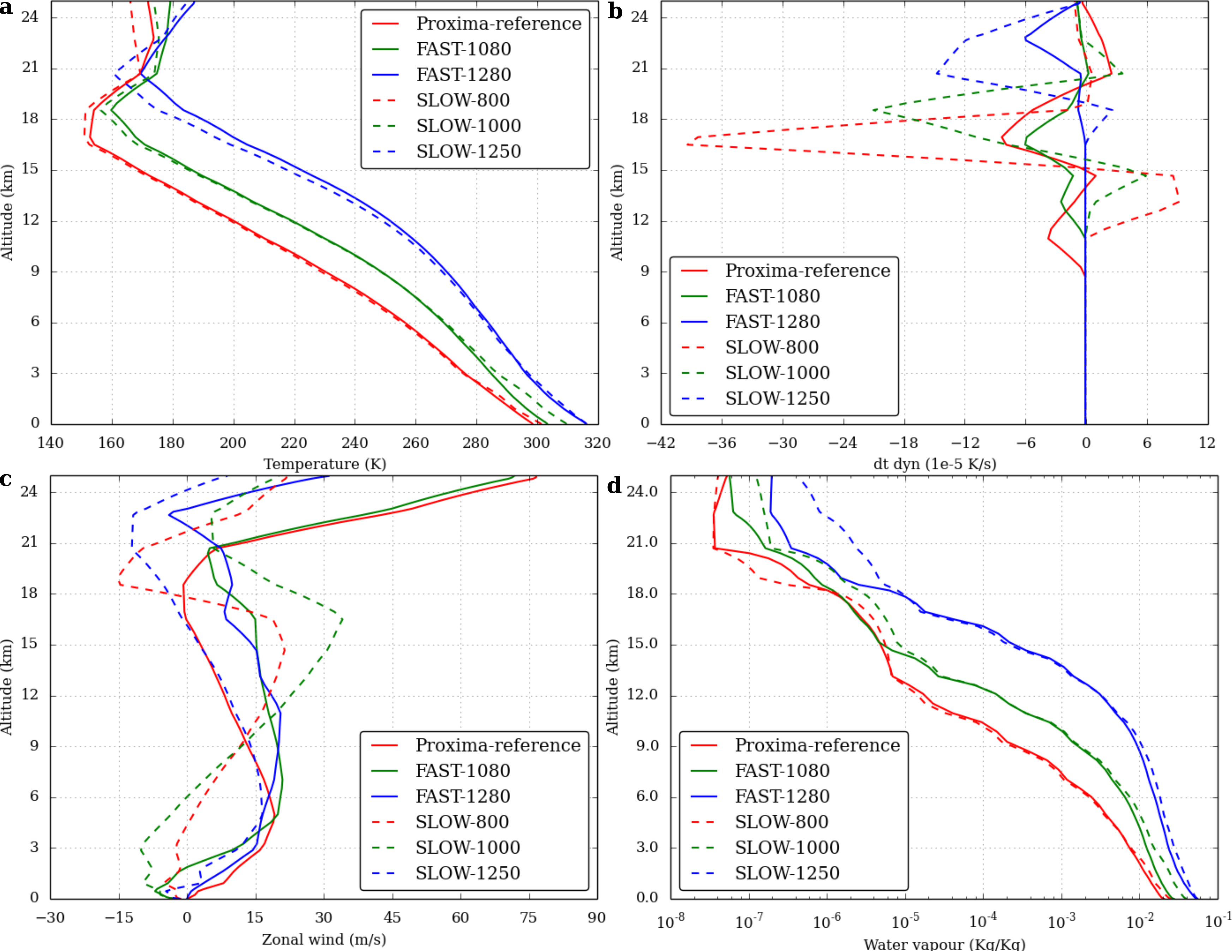}
 \caption{Initial vertical profile of the atmospheric temperature in K (a), advection heating rate in K/s (b), zonal wind in m/s (c) and water vapour in kg/kg of air (d) for the six case of stellar insolations and rotation rates considered in our study (see Table~\ref{T1}). These profiles are time-averaged GCM outputs at the maximum of column water cloud fraction (Fig~\ref{21}). The constant value of the large-scale from the surface to approximately 9~km is set to 10$^{-6}$~K~s$^{-1}$ (see text).} 
 \label{22}
\end{figure}
\end{center}

The position of the maximum cloudiness is close to the equator therefore no Coriolis effect is considered in the CRM runs. To take into account the heating/cooling from the large-scale circulation of the atmosphere, a heating rate profile extracted from GCM runs is added up to the temperature tendency from the physics package, in a similar way as in \cite{Lefe17,Lefe18}, called advection heating rate hereafter. These profiles are time-averaged and constant during the time of the simulations. On tidally-locked planets two circulation regimes prevail: equatorial super-rotating jet or substellar/anti-substellar circulation. The equilibrium depends on the the equatorial Rossby deformation radius \citep{Leco13}. Only the effect of the large-scale circulation on the temperature is considered in this study. For the planets considered in our study, i.e. an N$_2$-dominated atmosphere with a surface temperature around 300~K with a planetary radius close to the Earth, the atmospheric circulation falls into the two regimes. For a 11~days rotation period, the planetary atmosphere is in an equatorial super-rotating jet regime with planetary-scale waves interactions \citep{Show11,Hamm18}. For a 60~days rotation period, the equatorial Rossby deformation radius exceeds the planetary radius and therefore a planetary Rossby wave would be bounded by this radius, and thus the substellar/anti-substellar circulation dominates the large-scale circulation. This difference in circulation reverberates to differences in the advection heating rates, in particular between 15 and 25~km (see Fig~\ref{22}), and exhibits the greater efficiency of heat transport for slow rotators. A time-averaged GCM vertical profile of zonal and meridional winds is prescribed.

The general circulation therefore dominates the advection heating in the upper atmosphere while closer to the surface it is the convection. These rates are time-averaged from the GCM simulation, however to avoid any influence from the moist convective scheme used in the GCM physics package and let the resolved convection in the CRM free to equilibrate, the advection heating rate profile is set to a constant value of -1~$\times$~10$^{-6}$~K~s$^{-1}$ from the surface to approximately 9~km, representative of the cooling of the atmosphere at this altitude.
Fig~\ref{22}-a displays the initial temperature profile (K), Fig~\ref{22}-b the advection heating rate vertical profile (in 10$^{-5}$~K~s$^{-1}$ units), Fig~\ref{22}-c the initial zonal wind profile (m~s$^{-1}$) and Fig~\ref{22}-d the initial water vapor vertical profile (kg/kg of air). The atmosphere is assumed to be initially cloud-free. The stellar fluxes for the two rotation rates are slightly different to ensure a similar surface temperature which is expected to be the main driver of the convection and a key factor of the organisation of the convection. 3 sets of initial surface temperatures (calculated with the GCM simulations) are considered: 295~K, 305~K and 315~K at the substellar point. As shown in Fig~\ref{21} with the red dots, the extracted profile are not at the substellar point and so the surface temperatures differ over a few Kelvins from the reference surface temperature. The surface temperature is free to evolve during the simulation. The bottom boundary is an oceanic surface with a 0.07 albedo. The planetary and atmospheric parameters for the different cases considered are summed-up in Table~\ref{T1}.

The horizontal grid size is 1~km over 250~x~250~km to be able to revolve large convective cells, between two and three times the size of the LMD Generic GCM cells at the equator. We use a double periodic domain. The vertical domain extends from the surface to 25~km of altitude, about 900~Pa, to ensure that the convective plumes are able to reach their equilibrium levels. 
The number of vertical points is set to 80. A Rayleigh sponge layer is set at the top of the domain with a 1~km depth and a damping coefficient of 0.01~s$^{-1}$ to avoid spurious reflection of upward-propagating gravity waves \citep{Klem08}. The different cases are run over about 10 Earth days.

\begin{table}[!ht]
\begin{tabular}{|l|cccc|}
\hline
Parameter & \multicolumn{4}{ |c| }{Value} \\
\hline
Heat Capacity (J~K$^{-1}$) & \multicolumn{4}{ |c| }{1000} \\
Surface Pressure (Pa) & \multicolumn{4}{ |c| }{10$^5$} \\
Composition & \multicolumn{4}{ |c| }{N$_2$-dominated, CO$_2$ : 376 ppm, H$_2$O : variable amount (Fig~\ref{22})}  \\
\hline
Obliquity & \multicolumn{4}{ |c| }{0} \\
Eccentricity & \multicolumn{4}{ |c| }{0} \\
\hline
\hline
Case & Stellar Type & Incident flux & Rotation period & Gravity \\
  &  & (W~m$^{-2}$) & (days) & (m~s$^{-2}$) \\
\hline
Proxima b (reference) & M5.5 & 880 & 11 & 10.97 \\
FAST-1080 & M5.5 & 1080 & 11 & 10.97 \\
FAST-1280 & M5.5 & 1280 & 11 & 10.97 \\
SLOW-800 & M2 & 800 & 60 & 13.72 \\
SLOW-1000 & M2 & 1000 & 60 & 13.72 \\
SLOW-1250 & M2 & 1250 & 60 & 13.72 \\
\hline
\end{tabular}
\caption{Planetary and atmospheric parameters for the different simulations.}
\label{T1}
\end{table}

\section{Results for the reference simulation}
\label{Sec:ref}

We present in this section the results of our reference simulation, corresponding to the parameters (spectral insolation, mass, rotation rate, etc.) chosen to best reproduce the exoplanet Proxima b (corresponding to the first case in Table~\ref{T1}). 

\begin{center}
 \begin{figure}
 \includegraphics[width=17cm]{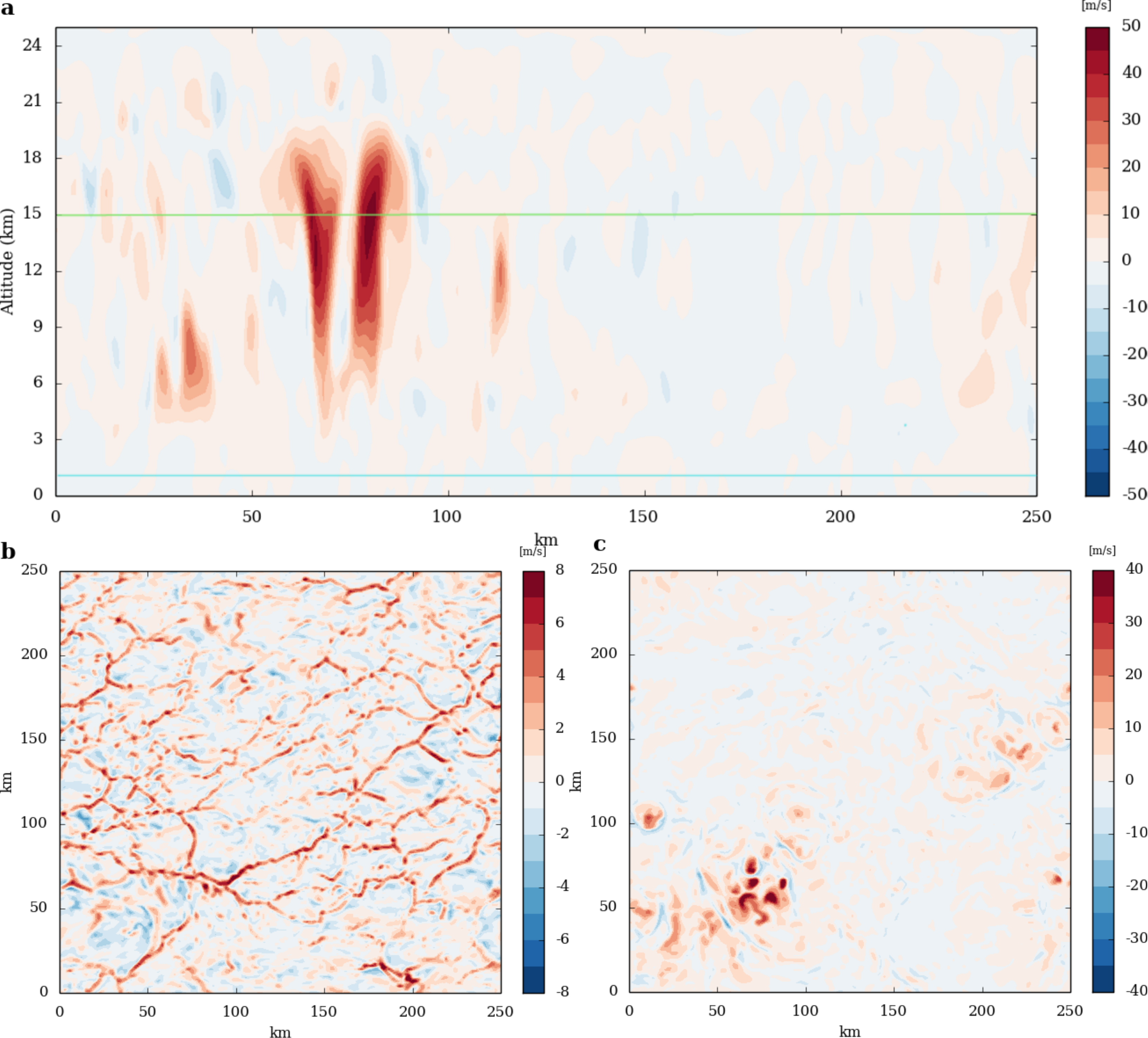}
  \caption{Screenshots after 30h of simulation of the vertical cross section (a) of the vertical wind (m~s$^{-1}$) and horizontal cross-section of the vertical wind (m~s$^{-1}$) at 1~km above the surface (b) and at 15~km above the surface (c) for the reference case. The cyan and green lines indicates the altitudes of the two horizontal cross-section, respectively 1 and 15~km. The vertical cross section is at y = 56~km.}
 \label{31}
\end{figure}
\end{center}

Fig~\ref{31} shows three snapshots of resulting vertical wind field (m~s~$^{-1}$) after about 30h of simulations: a vertical cross-section (a) and a horizontal cross-section at 1~km altitude (c) and 20~km altitude (c). Shallow convection is visible below 3~km with cells between 20 and 50~km of diameter and vertical wind reaching $\pm$ 8~m~s$^{-1}$, slightly above Earth shallow convection values at the same altitude \citep{Schu15,Gian16}. At 2.5~km, the typical vertical wind ranges between $\pm$ 1.5~m~s$^{-1}$, three times the values at this altitude in \cite{Serg20} study. This difference is due to the thickness of the shallow convection, thicker in the present study. The diameter of these cells is comparable to that of Earth shallow convection cells \citep{Atki96}. From 5~km to approximately 19~km (about 40~mbar), deep convection occurs with vertical wind reaching up to 40~m~s$^{-1}$. On Earth, deep convective plumes reach the tropopause at typically 15~km with vertical wind speed up to 20~m~s$^{-1}$ \citep{Zips80,Gian16}, in some extreme cases the plumes can reach 20~km altitude with vertical up to 50~m$^{-1}$ \citep{Dahu16}. The difference in low altitude vertical velocity with \cite{Serg20} is due to a difference of shallow convection depth in the present study, allowing higher vertical velocity. This difference in tropopause height engenders a higher convective available potential energy (CAPE) and leads to higher vertical velocity. Above 17~km and up to 20~km, tropopause-overshooting convection is visible. For comparison, above the United States only approximately 2~$\%$ of the overshooting convection reaches altitude 3~km above the tropopause \citep{Coon18}. In the absence of ozone heating in the model, the stratosphere lacks an inversion and is less stable, and could allow more penetration. At 15~km, the vertical wind is dominated by the large cluster (x~=~75 ; y~=~50)~km with a diameter of 50~km, but there are also several smaller plumes with 10~km diameter. The deep convective plumes reaching their top altitude will engender upward and horizontally propagating gravity waves transporting heat and momentum. Such strong convective activity is known to affect Earth's atmosphere through the quasi-biennal oscillation (QBO) atmospheric feature \citep{Dunk97}. Convectively generated gravity waves are thought to take part in similar QBO-like features in Giant-like and Brown Dwarfs atmospheres \citep{Show19}.

\begin{center}
 \begin{figure}
  \includegraphics[width=16cm]{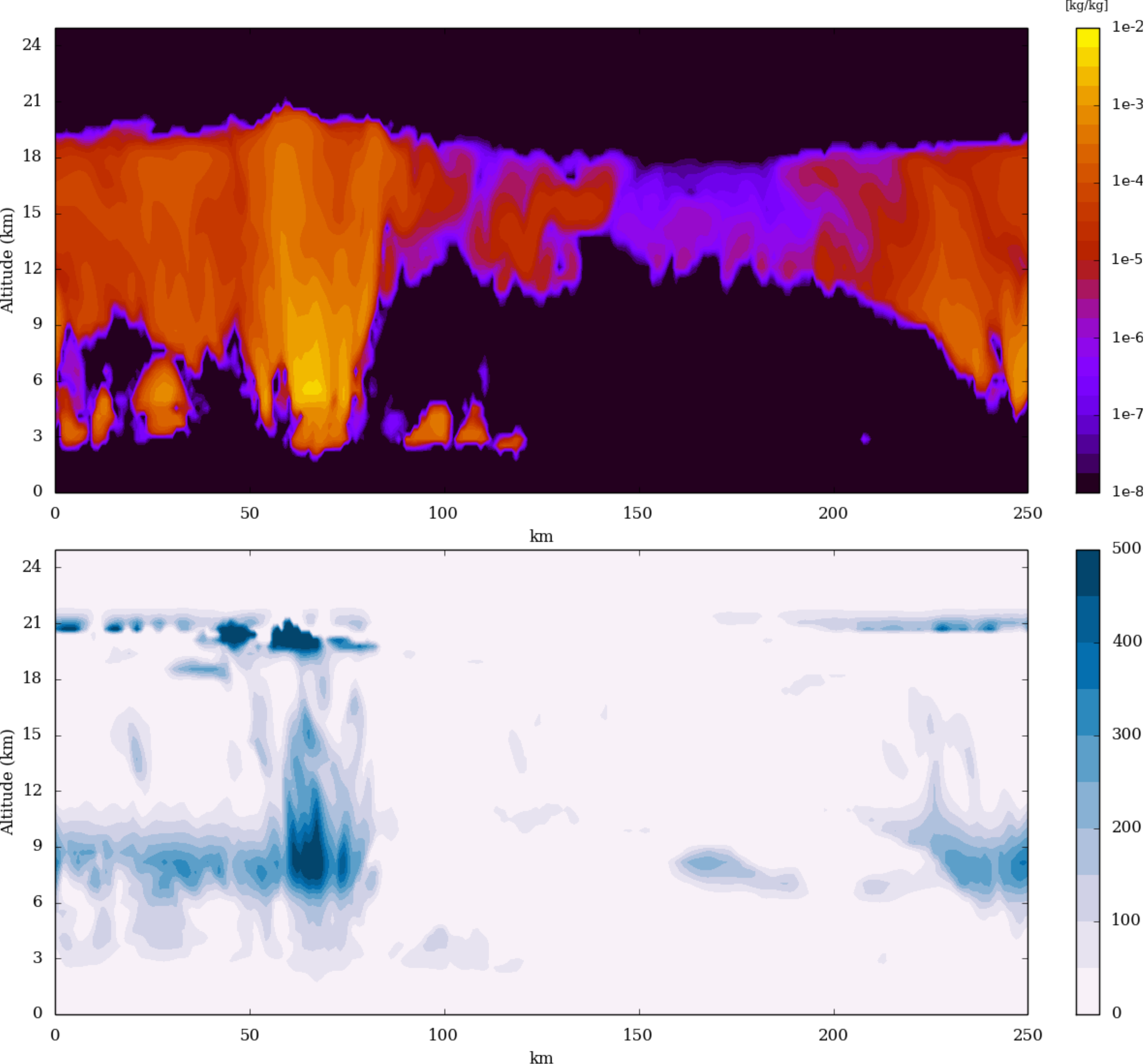}
 \caption{Screenshot of the the vertical cross section of the mixing ratio of water clouds (top) in kg/kg of air and of the water enrichment, i.e. the relative change, in $\%$ (bottom) for the reference case. For clarity the scale of values enrichment is limited at 500~$\%$ but it reach higher values. The vertical cross section is at y = 56~km.}
 \label{32}
\end{figure}
\end{center}

This deep convection transports water vapour from 4~km up to the tropopause, leading to water condensation and thus cloud formation. Fig~\ref{32} shows a screenshot of a vertical cross section of the water clouds abundance (top) and a vertical cross-section of the water vapour enrichment (bottom), i.e. the relative change regards to the mean water vapor vertical profile.

\begin{center}
 \begin{figure}[!htp]
 \centering
  \includegraphics[width=11cm]{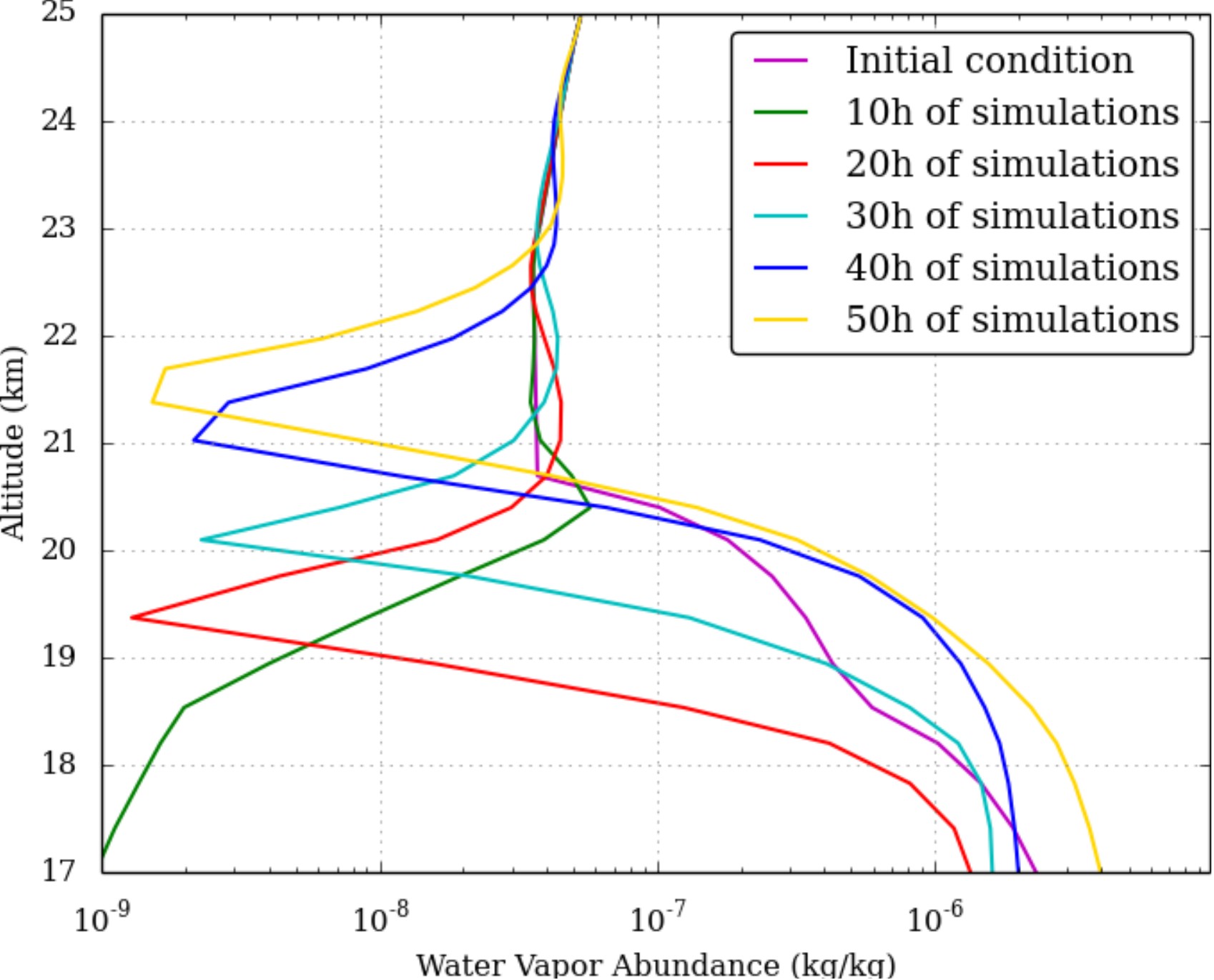} 
 \caption{Vertical profile of domain-averaged water vapor content (kg/kg) in the stratosphere at initial condition, after 10~h, 20~h, 30~h, 40~h and 50~h of simulations for the reference case.}
 \label{35}
\end{figure}
\end{center}

Between 12 (160~mbar) and 18~km, there is the presence of a complete cloud cover, as obtained in past GCMs study \cite{Yang13}.
The enrichment of water vapour at x~$\sim$~75~km is related to the large vertical plumes visible in Fig~\ref{31}a, reaching up to 500~$\%$ enhancement compared to the mean water vapour abundance at this same altitude. The maximum of this enrichment is above 20~km due to convection overshooting with a value as high as 10 000~$\%$ 
The water is then advected horizontally by winds and visible between 200 and 250~km in the x axis. On Earth the water vapour enrichment at 100~hPa due to overshooting convection can reach up to 300~$\%$ \citep{Herm17}. Fig~\ref{35} shows the evolution with time of the stratospheric water vapour content (kg/kg of air). Soon after initialization, there is an abrupt decrease of water vapor due to the water vapour condensation, after the organization of convection the water vapor is transported by the large plume resulting to enrichment visible below 21~km with a maximum of increase by almost a factor 4 around 19~km. This water vapor advected by the convection from the lower atmosphere to the stratosphere is then advected by wind and can condense elsewhere on the planet. Between 21~km and 23~km the atmosphere becomes drier.

\begin{center}
 \begin{figure}[!h]
 \centering
 \includegraphics[width=10cm]{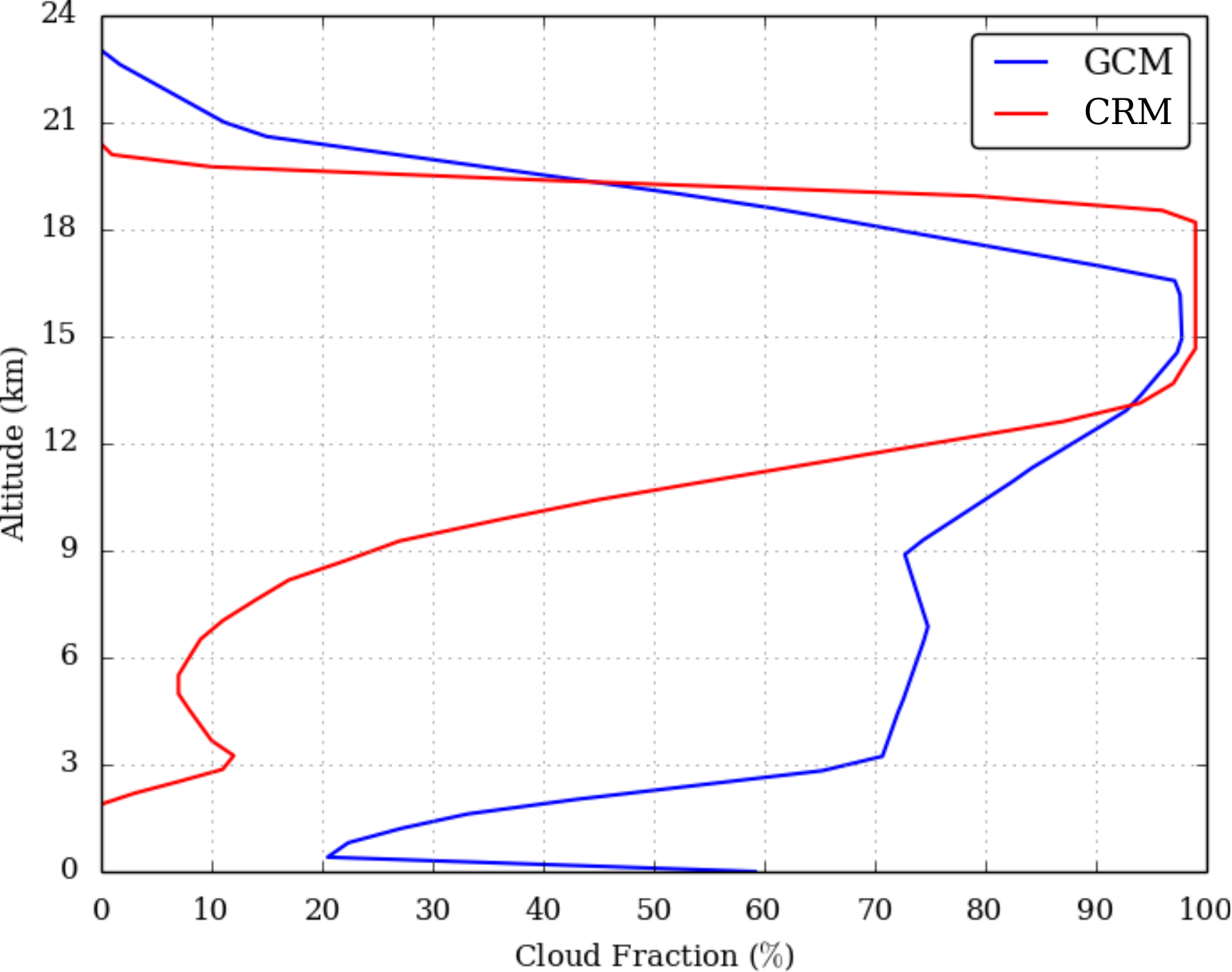}
 \caption{Comparison between the domain-averaged vertical profile of the cloud fraction obtained with the CRM (red) for the reference case and the time average vertical profile of the cloud fraction from the LMD Generic GCM (blue) at the chosen position (see Fig~\ref{21}).}
 \label{33}
\end{figure}
\end{center}

Fig~\ref{33} displays the cloud fraction for the Proxima standard case.
The equivalent of Earth stratocumulus clouds are visible below 5~km and covers about a tenth of the surface, compared to 20$\%$ for stratocumulus above tropics ocean \citep{Eas11}. Between 12 and 19~km the cloud cover increase up to almost 100~$\%$ as predicted by GCM modelling \citep{Yang13}. Surface rain over the domain is equal to 15 mm/day. Under the deep convective plume, the surface rain reaches 120~mm/day.

The cloud cover is much lower in this CRM study compared to the GCM, especially for the low altitude clouds which will greatly impact the radiation reaching the surface. The cloud fraction in the GCM is non-zero from the surface, with almost 60~$\%$ to about 40~$\%$ at 2~km, whereas for the CRM model the cloud fraction is non-zero from 2~km, where the shallow convection ends. At this altitude the GCM cloud fraction is larger than 45~$\%$. This substantial discrepancy in cloud cover could explain the difference in albedo (discussed hereafter) between the CRM and GCM models and shows the way to deal with the shallow convection.
The high cloud layer is quite comparable in depth but at different altitude, 3~km higher for the CRM. 

\begin{center}
 \begin{figure}[!h]
 \includegraphics[width=17cm]{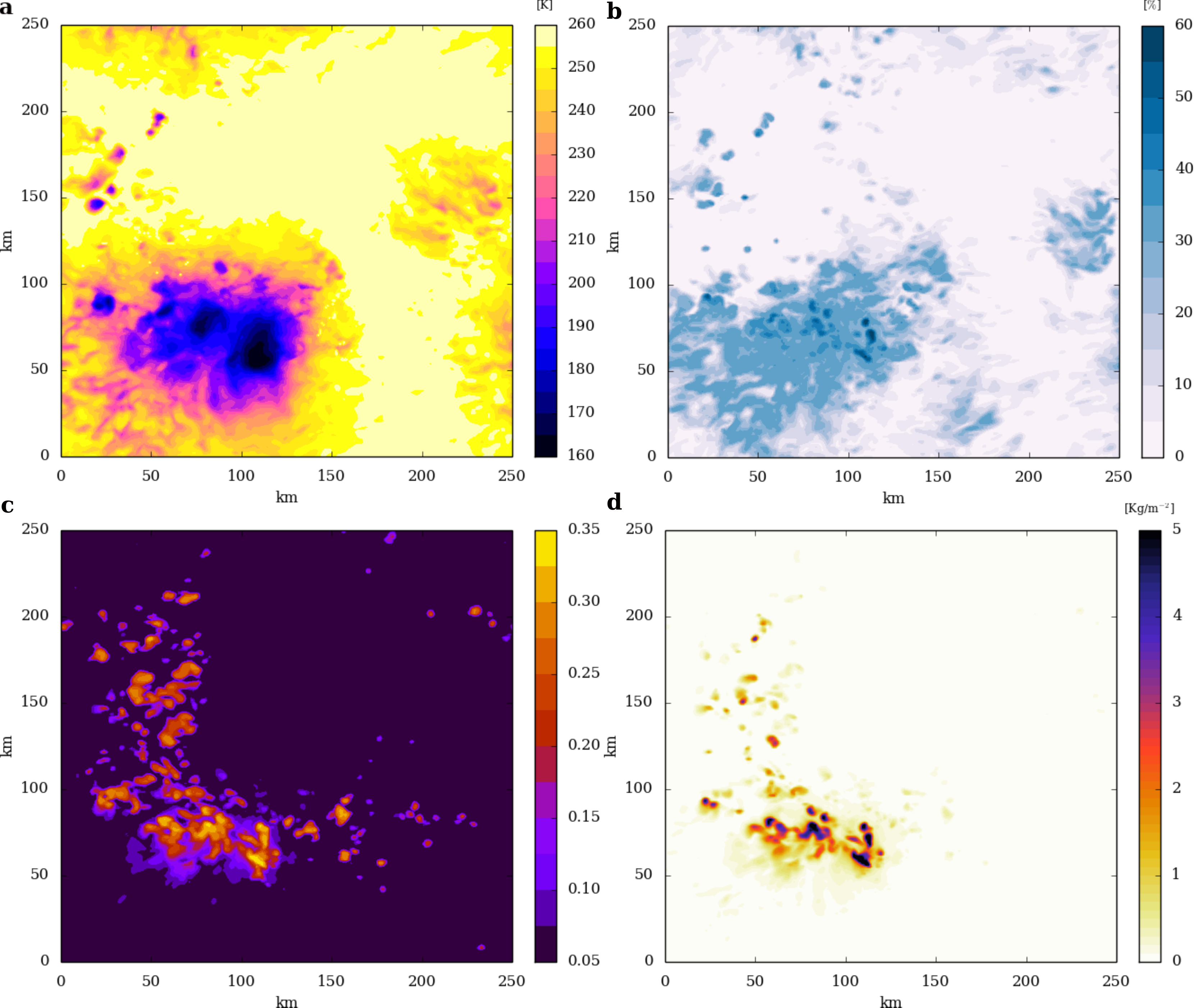}
 \caption{Screenshot of the OLR brightness temperature in K (a), the relative humidity at 8~km (b), the bond albedo (c) and vertically integrated water clouds abundance in Kg~m$^{-2}$ (d) for an incident stellar flux of 880~W~m$^{-2}$ and a 11~days rotation period.}
 \label{34}
\end{figure}
\end{center}

These clouds impact the fraction of shortwave radiation able to reach the surface but also the longwave infrared thermal emission. Fig~\ref{34}-a shows the outgoing longwave radiation (OLR) in brightness temperature units (K). The large cluster with a diameter of 150~km -- where the OLR reaches a minimum value -- corresponds to the vertical plume visible at x~$\sim$~75~km in Fig~\ref{31}-a. The brightness temperature drops to about 160~K in that structure, corresponding to an altitude of 16~km (Fig~\ref{22}a). Outside of the cluster the OLR can reach values up to 260~K, equivalent of an altitude of 6~km above the surface. This structure resemble convection aggregation \citep{Wing14} but is transient, with a lifetime of about 3 hours, typical for a convective structure of that size \citep{Roca17}. After it collapses, an other structure with similar size is emerging at a different location of the 250~x~250km grid. Circular patterns of a few tens of kilometers are also visible. They are associated with the presence of smaller convective plumes. Fig~\ref{34}-b shows a horizontal cross section of the relative humidity at 8~km. The deep convective cluster is drying the non-convection area, acting like self-aggregation on Earth \citep{Tobi12}.

Fig~\ref{34}-c shows the corresponding Bond albedo map and Fig~\ref{34}-d shows the corresponding vertically integrated water cloud (Kg~m$^{-2}$). Albedo can reach values from 0.05 in cloud-free regions up to 0.34 in the regions corresponding with the positions of high altitude convective plumes, where there is the largest amount of clouds (reaching up to 5~kg~m$^{-2}$ in the column). The low altitude stratocumulus clouds contribute for a significant part of the high albedo values as visible at x~=~160~km with less than 1~Kg~m$^{-2}$ in the column. The domain averaged value of the Bond albedo is 0.06 whereas the GCM Bond albedo is 0.32. This is explained firstly by the fact that the cloud coverage ($\sim$~10$\%$ for low altitude stratocumulus clouds) is relatively low; and secondly by the fact that the cloud-free regions have a very low bond albedo. The latter stems from the fact that the albedo of the ocean surface is low (0.07), that the Rayleigh scattering is inefficient around a cool star such as Proxima Cen, and that the water vapour-enriched atmosphere absorbs efficiently near-infrared emission from the star. A similar reduction of the bond albedo is also observed in the 1-D cloud-free simulations of \cite{Kopp13} (see Fig.~6a, calculated for a surface albedo of 0.3).

The maximum bond albedo value in the CRM, corresponding to the large convective cluster, is comparable to the mean GCM bond albedo at the substellar point. This substantial discrepancy could be explained by the significant difference in low altitude clouds fraction between the CRM and GCM models and difference of shallow convection. This value of albedo can be compared to the TOGA-COARE case with Sun spectra in the Appendix~\ref{App}, where the albedo inside the deep convective plume is around 0.6 whereas the albedo is about 0.1 in the cloud-free region. The low value of the overall albedo is much smaller than previous GCM previsions \citep{Yang13,Yang19}, which reflects the fact that outside deep convective layers where the albedo is close to previous GCM studies, the areas of low cloud fraction have low cloud albedo and therefore aggregation of convection, even transient like in the study, can have a strong effect on the overall albedo due to the low mass star spectra. 

\begin{center}
 \begin{figure}[!h]
 \centering
 \includegraphics[width=16cm]{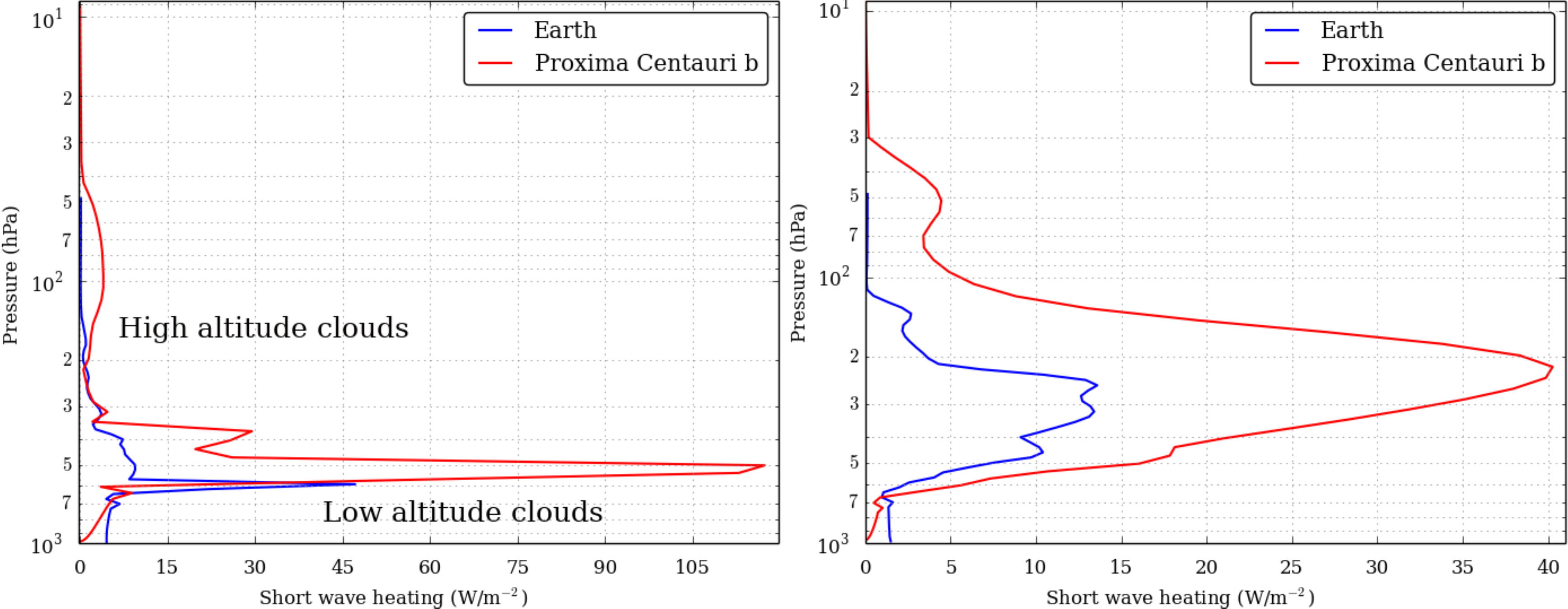}
 \caption{Vertical profiles of incident stellar heating (in W~m$^{-2}$, per atmospheric layer). The red lines are for Proxima Centauri b reference case and the blue lines are for the TOGA-COARE case. The left figure is for the case of the presence of both high altitude clouds and low altitude clouds and the right panel is for the case of deep convection.}
 \label{38}
\end{figure}
\end{center}

In addition to its synchronous rotation, another specificity of Proxima Centauri b compared to Earth is the spectra of its host star. Proxima Centauri is an M5.5 type star \citep{Bess91} with an effective temperature of 3050~K with a maximum of the emission spectra around 1~$\mu$m \citep{Pavl17}. This shift towards the infrared is expected to impact the absorption of the stellar spectra by the atmosphere, especially by water. Fig~\ref{38} shows vertical profiles of the stellar heating (W~m$^{-2}$) for two water cloud features for both the TOGA-COARE case (blue line) with sun spectra and the Proxima Centauri b reference case with M-star spectra (red line).

The two water cloud features displayed are high altitude and low altitude clouds and deep convection. On the left figure for Earth, high clouds absorb solar heating around 1.5~$\times$~10$^2$~hPa while the absorption by the low altitude clouds occurs at 6~$\times$~10$^2$~hPa. For Proxima Centauri b, high clouds absorb solar heating between 2~$\times$~10$^2$~hPa and 4 and 7~$\times$~10$^1$~hPa with a maximum at 1~$\times$~10$^2$~hPa and the absorption of the low altitude clouds occurs at 5~$\times$~10$^2$~hPa for the reference case. The short wave heating by the low clouds on Proxima Centauri b is almost three times stronger that Earth, shielding more efficiently the surface from the incoming stellar radiation. For the deep convection case (right figure), for both Proxima Centauri b and Earth the short wave heating is absorb on a larger vertical scale, the heating is also stronger for Proxima Centauri b, more than three times the one on Earth. 
A larger part of the energy coming from the host star is absorbed by the atmosphere gas and clouds, and therefore even with a lower overall albedo, the energy reaching the surface is lower for a planet orbiting of M-star than a Sun-like star (for an Earth-like atmosphere).

\section{Impact of the incident flux}
\label{Sec:flu}

\begin{center}
 \begin{figure}
  \includegraphics[height=10cm,width=16.5cm]{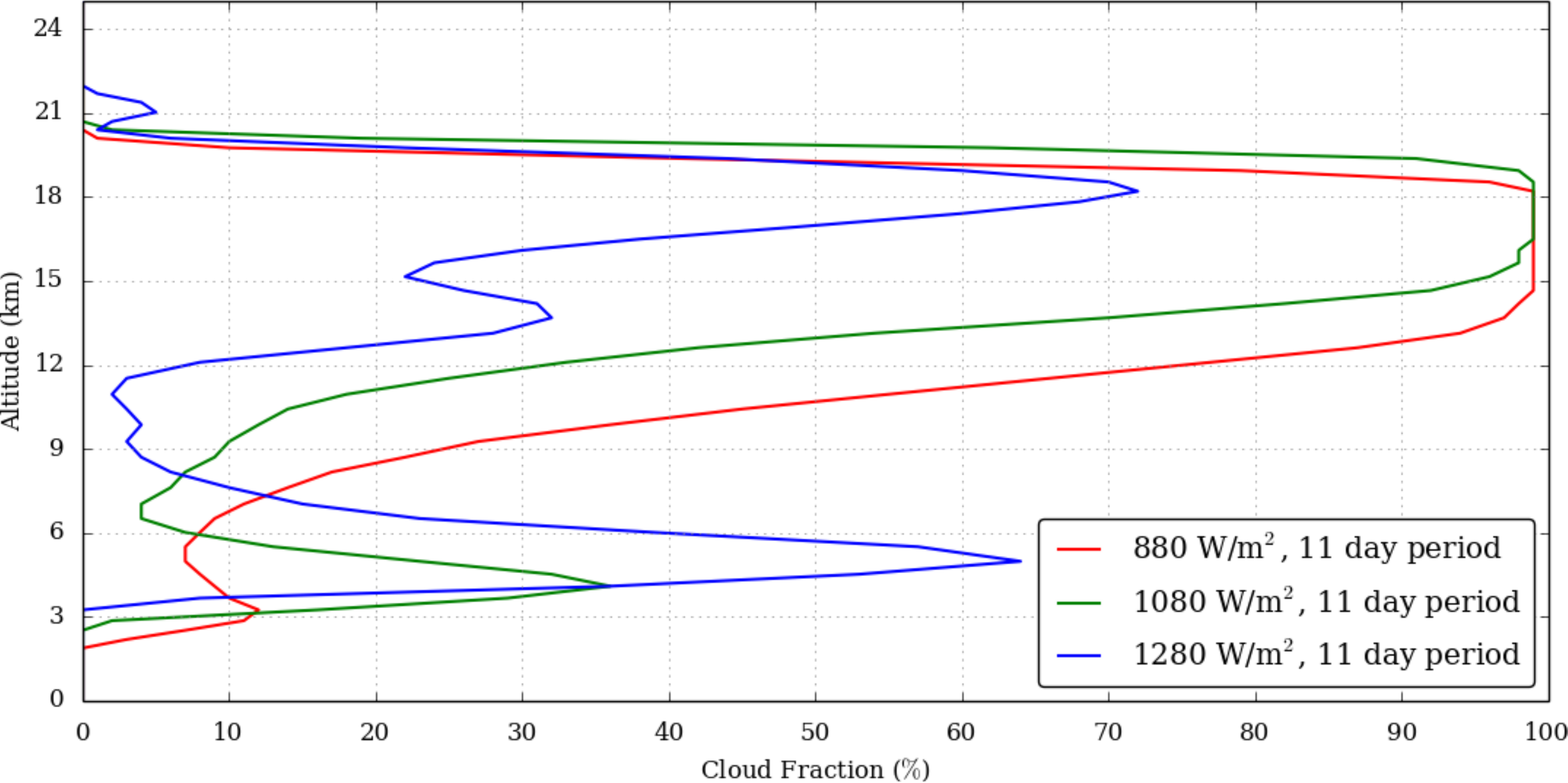}
 \caption{Comparison of the domain-averaged cloud fraction for the reference case (Top), FAST-1080 (Middle) and FAST-1280 (Bottom) cases.}
 \label{41}
\end{figure}
\end{center}

Fig~\ref{41} shows the comparison of the cloud coverage for the reference case with the FAST-1080 and FAST-1280 cases (all with a 11~days rotation period). In all the three cases, there is a high-altitude cloud layer. The altitude of this layer is determined by the thermal structure. The tropopause increases with the stellar flux, with a value of 17~km (50~hPa) for the standard case, 19~km (32~hPa) for the FAST-1080 and 21~km (20~hPa) for the FAST-1280 case (see Fig~\ref{22}a). The thickness of this high-altitude cloud layer decreases with the increase of the incident flux. This inverted trend is due to the low altitude clouds. The increasing stellar flux engender a thicker shallow convection, from around 3.0~km for the reference case to 5.6~km for the FAST-1280 case, leading to more extented vertical mixing of the water vapor in the shalow convection. Domain-averaged value for the standard case is 8~x~10$^{-3}$~kg/kg, 1.1~x~10$^{-2}$~kg/kg for the FAST-1080 case and 2.3~x~10$^{-2}$~kg/kg for the FAST-1280 case. This increase in water vapor by the shallow convection triggers more condensation, leading to an increase of the low-altitude clouds coverage, reaching up to 60$\%$ for the FAST-1280 case. 

\begin{center}
 \begin{figure}
  \includegraphics[width=17cm]{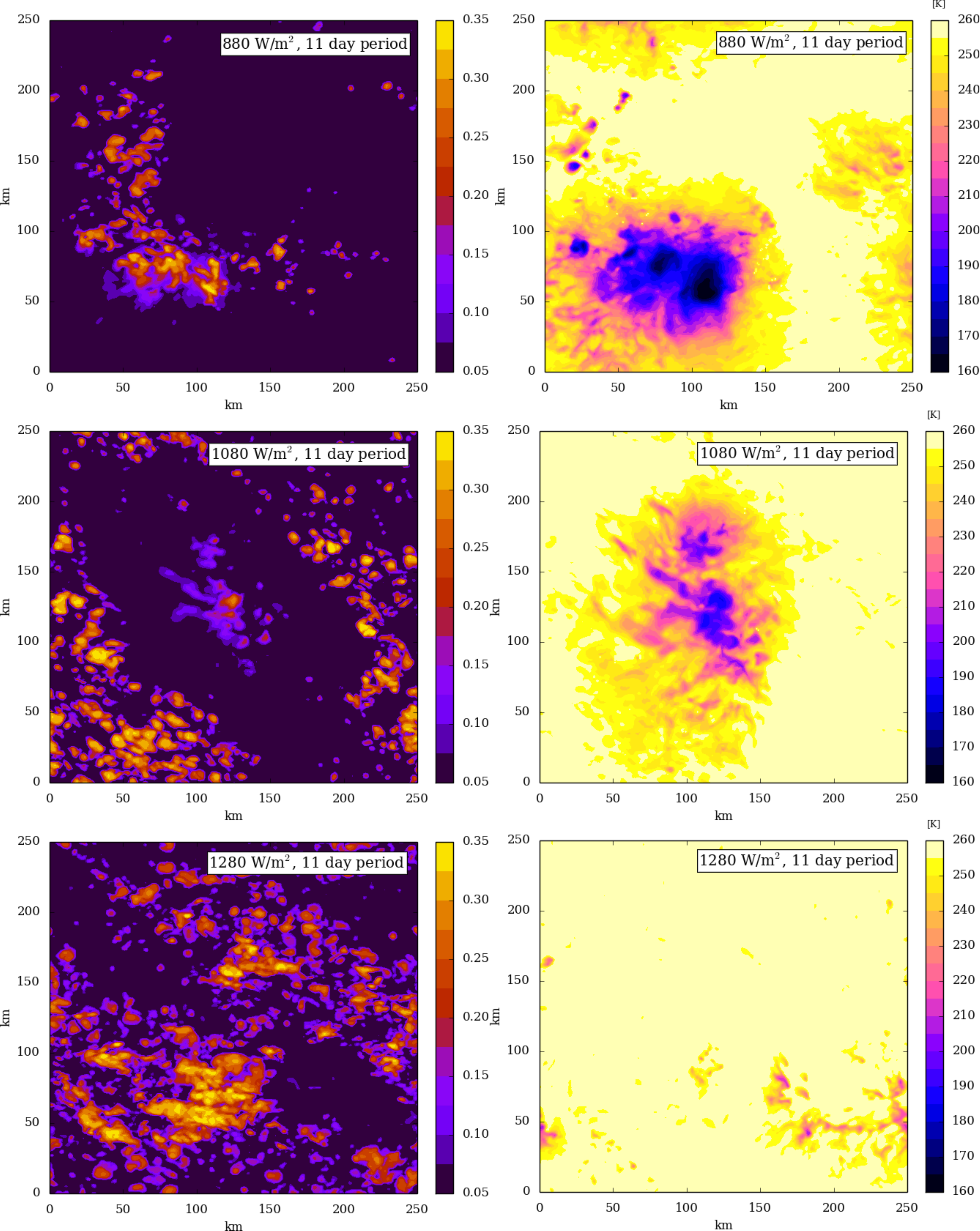}
 \caption{Snaphots of albedo (Left) and OLR brightness temperature in K (Right) for Proxima b (Top), FAST-1080 (Middle) and FAST-1280 (Bottom) cases.}
 \label{42}
\end{figure}
\end{center}

Fig~\ref{42} shows the atmospheric albedo (left) and the OLR brightness temperature (K) for the reference case (Top), FAST-1080 (Middle) and FAST-1280 (Bottom) cases. One noticeable behaviour is the increase of both the atmospheric albedo and the OLR with the increase of the stellar flux. The structure of the albedo map for the standard case is dominated by the high cloud layer visible in the OLR map. 
For the FAST-1080 case, the contribution of high-altitude clouds is visible in the center of the map but the low-altitude cloud layer is dominating with small patchy structures. The contribution of low-altitude clouds is even more important for the FAST-1280 case where the contribution of the high-altitude clouds, visible on the OLR map, is barely noticeable.

With the increase of the stellar flux, the thermal structure of the atmosphere changes with an increase of the tropopause height as well as an increase of the atmospheric water vapor mixing ratio. This increase of stellar flux induces a thicker shallow convection leading to an increase of the low altitude cloud layer coverage. This increase of the low altitude clouds engender an increase of the atmospheric albedo and a stronger short wave cloud radiative feedback. This temperature trend is consistent with anvil cloud fraction SST dependence on Earth \citep{Bony16}, where an increase of the SST reduces the cloud fraction. 

Simulations with initial surface temperature around 350~K, corresponding to incoming stellar flux of 1500 and 1610 W~m$m^{-2}$ for the FAST and SLOW cases respectively, was conducted but not shown here. The abundance of water vapor in the atmosphere reaches values superior to 20\%, going over the dilute regime, therefore a condensation of water vapor would lead to a significant pressure change not taken into account by the model. To be able to study such temperatures range changes need to be carried out to take into the non-dilute regime.

\section{Impact of the rotation period}
\label{Sec:rot}

The rotation period has an important impact on the circulation of a planetary atmosphere. The strength of the Coriolis force dictates the extend of the thermally direct latitudinal circulations and the surface temperature distribution, which drives atmospheric circulation \citep{Pier10}. From Fig~\ref{22}-a, we observe that the thermal structures and water vapour vertical distributions are similar for the slow and fast rotation cases (for a given surface temperature), whereas the advection heating rates have different behaviour (Fig~\ref{22}-b). For the SLOW cases, the profile of advection heating rates for the 3 cases of solar insolation have similar behaviour with positive values region between 12 and 18~km and above a strong negative values region between 16 and 21~km. On the other hand, for the FAST cases the advection heating rates have only negative below 21~km.

\begin{center}
 \begin{figure}
  \includegraphics[height=10cm,width=16.5cm]{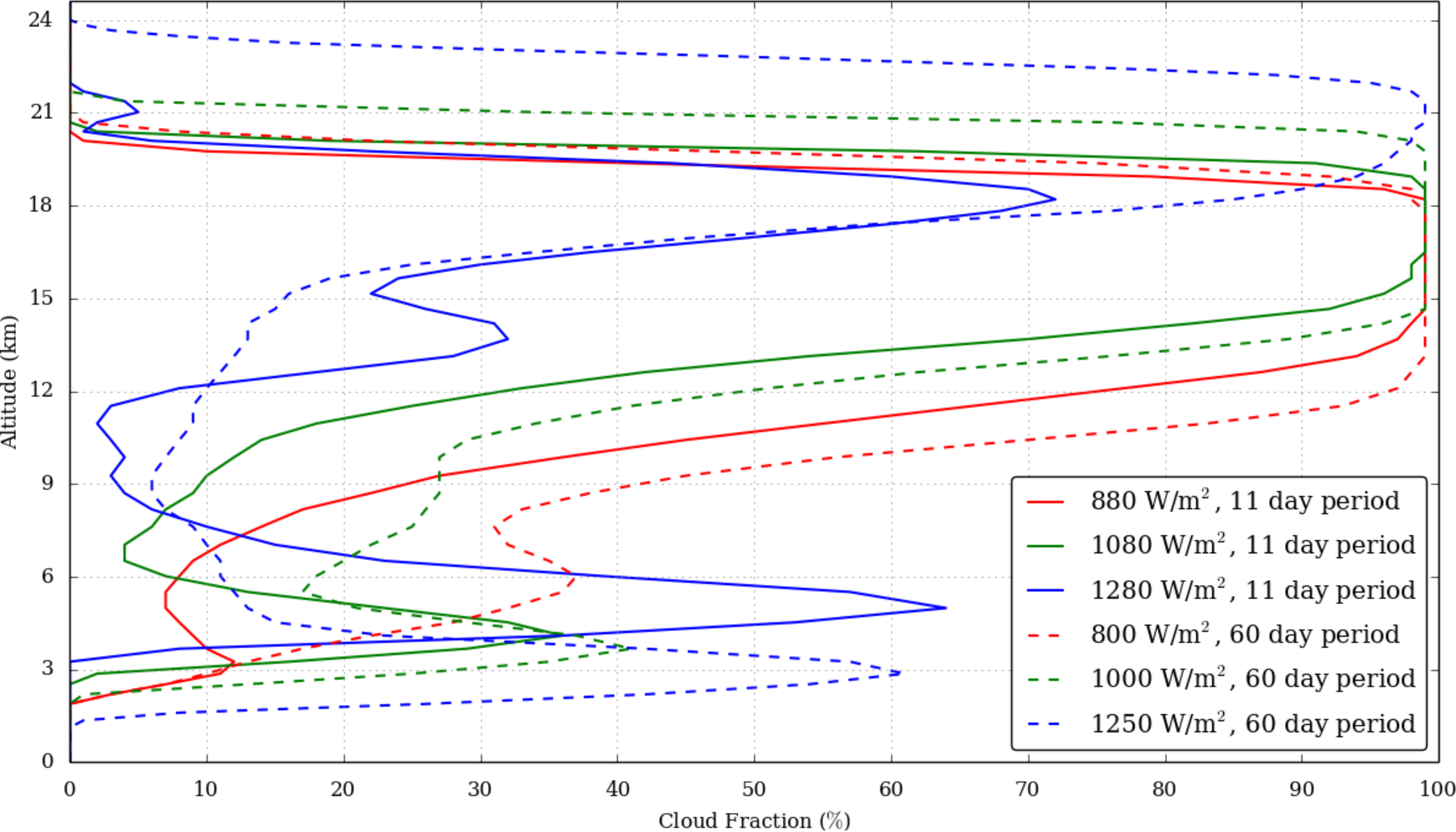}
 \caption{Comparison of the domain-averaged cloud fraction for the SLOW-800 (Top), SLOW-1000 (Middle) and SLOW-1250 (Bottom) cases.}
 \label{51}
\end{figure}
\end{center}

Fig~\ref{51} shows the comparison of the cloud coverage for the 6 cases studied in our work: the reference case, FAST-1080 and FAST-1280 with a 11~days rotation period and SLOW-800, SLOW-1000 and FAST-1250 with a 60~days rotation period. The height of the tropopause is very similar for the two rotation periods but the advection heating rates are quite different with stronger values for the 60~days rotation rate.
For the 6 cases, there is a high-altitude cloud layer thinner with increasing stellar flux as described in the previous section. The depth of this cloud layer also increases with the rotation rate. The high-altitude cloud layer is about 1.5~km thicker for the SLOW-800 compared to the reference case, 2~km thicker for the SLOW-1000 case compared to the FAST-1080. For the SLOW-1250, this layer is not only thicker but it reaches higher cloud fraction values. The low-altitude cloud coverage is very similar for the four highest incident stellar flux cases whereas for the two lowest there is a significant difference with the rotation period. The low altitude cloud coverage of the SLOW-800 case is almost four times the coverage of the reference case. 
The altitude of the lower cloud deck is varying with the incident flux but also with the rotation period. For the 60~days cases, the top of the shallow convective layer is lower, especially at high stellar flux. This effect increases the albedo and can be  attribute to the thicker high altitude cloud layer absorbing and reflecting more incident stellar flux.

\begin{center}
 \begin{figure}
  \includegraphics[width=17cm]{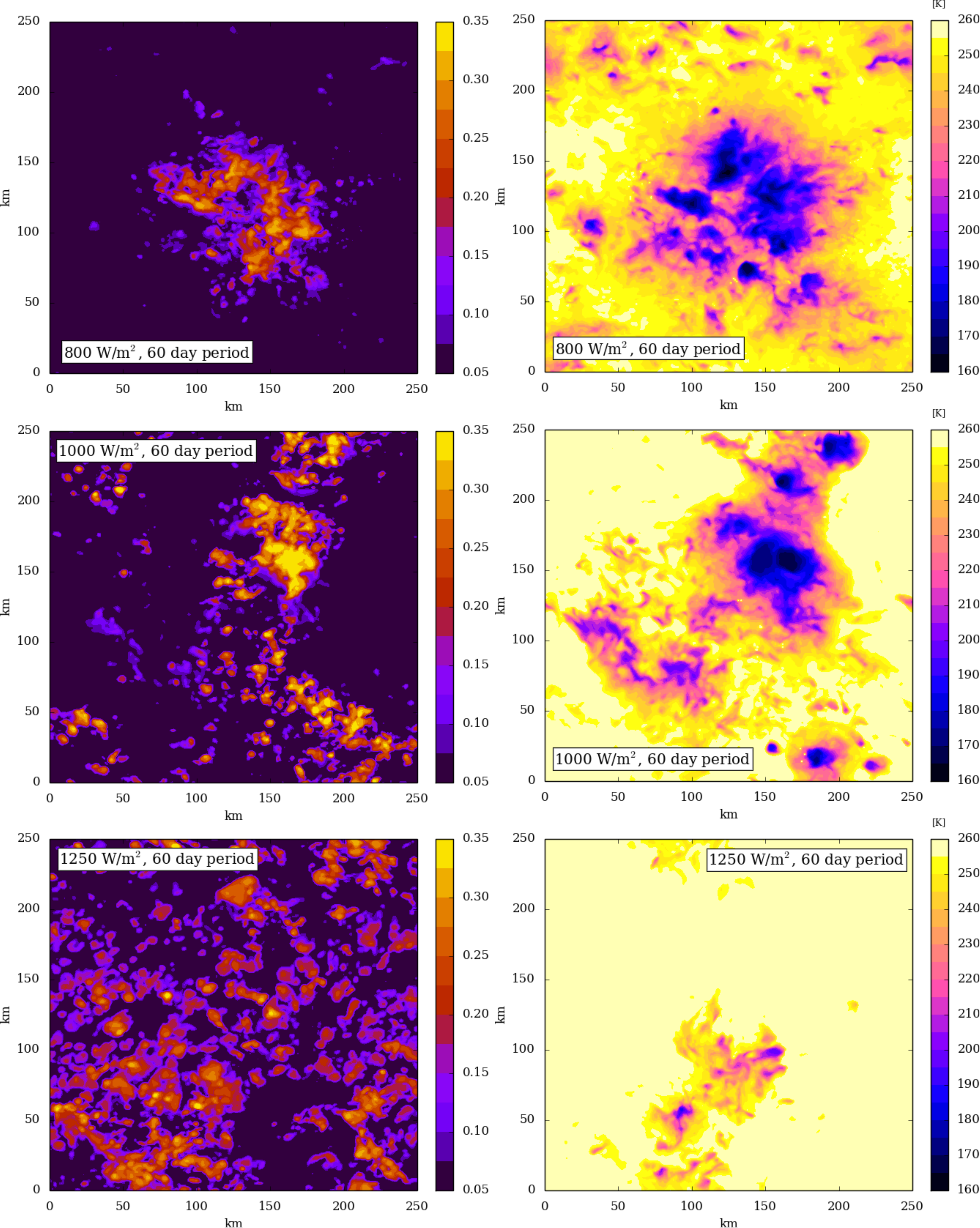}
 \caption{Snaphots of albedo (Left) and OLR brightness temperature in K (Right) for SLOW-800 (Top), SLOW-1000 (Middle) and SLOW-1250 (Bottom) cases.}
 \label{52}
\end{figure}
\end{center}

Fig~\ref{52} shows the albedo (left) and the OLR brightness temperature for SLOW-1000 (Top), SLOW-1000 (Middle) and SLOW-1250 (Bottom) cases. Compared to Figure~\ref{42}, the OLR brightness temperature is lower for the 60~days rotation period than for the similar 11~days rotation case. For example, the large-scale structure visible in the Figure~\ref{52} (bottom right) for the SLOW-1250 has a stronger OLR flux, i.e. a smaller brightness temperature, than the large-scale structure visible in the Figure~\ref{42} (bottom right) for the FAST-1280 case. This is due to the fact that the high-altitude clouds is more developed for the 60~days rotation period and that the updrafts are stronger. The increase of low clouds albedo with the with increase of surface temperature is also noticeable with the a 60-days rotation rate as for a 11-days rate. 

In summary, the increase of the rotation period engenders a change of circulation regime leading to a different behaviour of the advection heating rate in altitude. This increase of the high altitude cloud layer thickness is linked to the vertical structure of the profile of advection heating rates between 12 and 21~km. This structure with positive values followed by negative of heating rates will trigger additional convective activity in this region. The increased thickness of the altitude cloud layer correspond to the region where this triggering takes place, between 12-19~km for the SLOW-800 case, 15-20~km for the SLOW-1000 case and 18-24~km for the SLOW-1250 case. This triggering will increase the convective activity and the vertical transport of water vapor in this region, and therefore engender a thicker cloud layer. With this change in high-altitude clouds, a decrease of shallow convective layer is occurring leading to an increase of the Bond albedo and cloud feedback. 

\section{Discussions}
\label{Disc}

\begin{center}
 \begin{figure}
 \includegraphics[width=12cm]{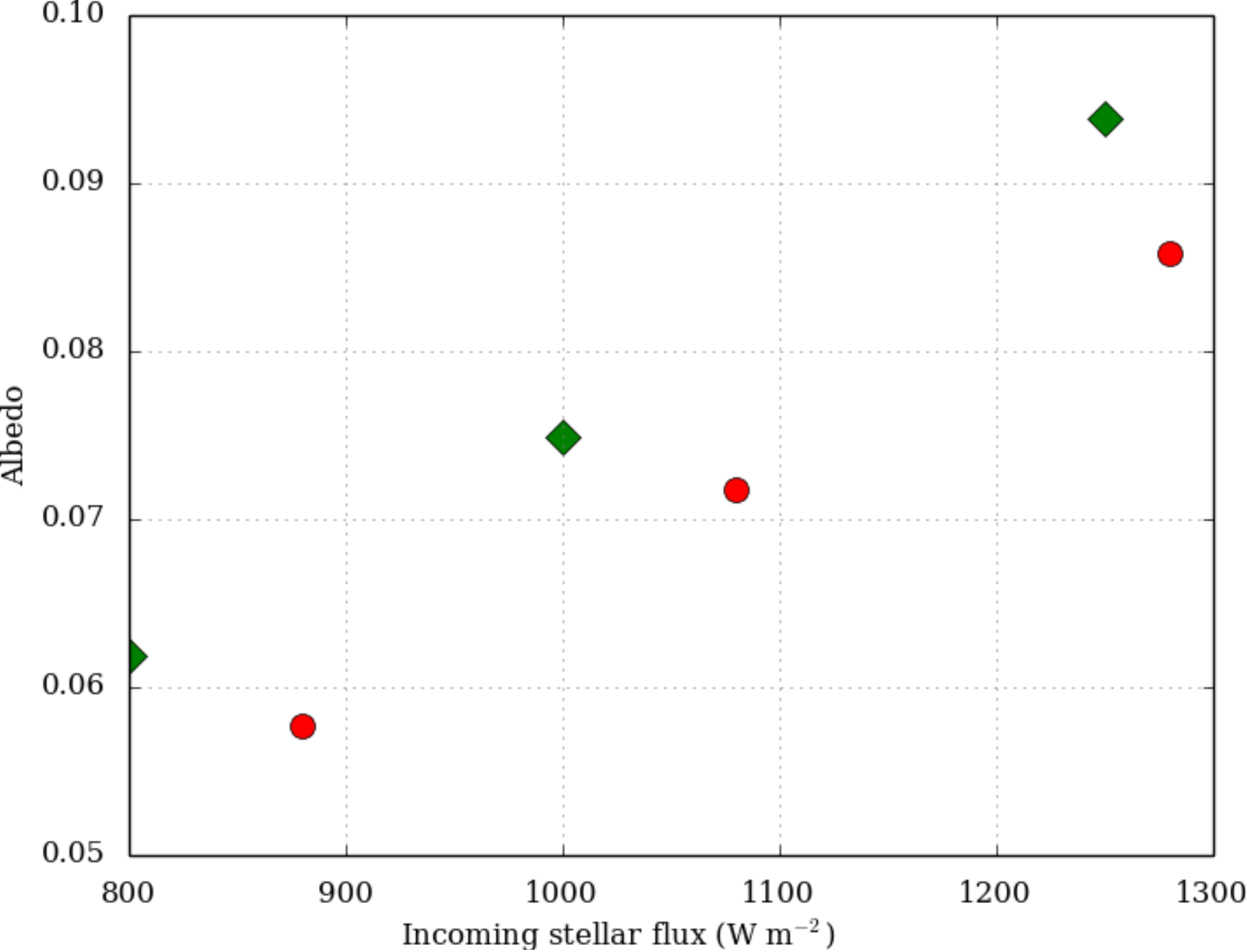}
 \caption{Domain-averaged bond albedo as a function of the incident stellar flux (in W~m$^{-2}$) for the CRM. Diamonds are for SLOW cases and circles for the FAST cases.}
 \label{61}
\end{figure}
\end{center}

Fig~\ref{61} shows the comparison of the domain-averaged Bond albedo for the 6 cases as a function of the incident stellar flux. The trends discussed in Sections~\ref{Sec:flu} and \ref{Sec:rot} are visible. The bond albedo is increasing of about 20\% with the increasing flux and is increasing of between 5 and 10\% with increasing rotation period. The study of \cite{Yang13} found a similar trend in the increase of bond albedo with insolation: between stellar fluxes of 1000 and 1200~W~m$^2$, the bond albedo increases by 7\%, although with much higher values around 0.45. 
Using a GCM with a high-resolution zoom in the substellar region, \cite{Serg20} studied the convection on tidally-locked Earth-like planets. A notable discrepancy is the fact that the albedo inside the high-resolution area is higher than with the low-resolution GCM in \cite{Serg20}, whereas in this study the albedo of the CRM is lower than in the GCM. The main noticeable difference between the two models is that in \cite{Serg20}, the large-scale forcing is given by the GCM fields (temperature, water content, winds, ...) in real-time whereas in this study the large-scale forcing is represented by a single time averaged vertical profile in temperature fixed in time and space. This difference in handling the large-scale forcing could lead to a difference of the shallow convection depth and therefore a difference in albedo.
The aggregation of convection showed in Section~\ref{Sec:ref}, and not seen in \cite{Serg20} study, is also one factor of this albedo discrepancy.
These differences between the two models call for more convection resolving modelling in such atmospheric environment leading to an intercomparison, possibly in the same way the THAI GCM intercomparison for exoplanets \citep{Fauchez:2019gmd} was carried out, in order to improve the understanding of convection processes in exoplanetary environments. 

The model does not make the difference between liquid water cloud and water ice cloud.
The impact of the general circulation is represented in the model by a constant heating rate profiles, a water vapor tendency profile could be imposed to the domain to take into account the water vapor advection near the substellar point. The temperature impact of the general circulation is represented with a constant tendency profile close to the surface, a more realistic approach with a time-dependent profile could improve the physical representation of the advection near the substellar point. 
Another improvement of the model will be to take into account ocean circulation. \cite{Delg19} showed the impact of such circulation on the SST and the clouds self-aggregation \citep{Mulle2020} and how SST anomalies can favor the aggregation of convection \citep{Sham20}. The oceanic circulation is also believed to have a significant impact on the atmospheric circulation, with a planetary heat transport dominated by the ocean at moderate incident stellar fluxes and by the atmosphere near the inner edge of the habitable zone \cite{Yang19b}. This is assuming an oceanic surface, the case of continental surface also needs to be carried out. 

\section{Conclusion}
\label{Conc}

We presented the LMD Generic CRM coupled to realistic radiative transfer and a non-Earth based cloud microphysics model. This versatile model is able to resolve realistic Earth tropics convection as well as to perform in extreme insolation conditions such as those experienced on tidally-locked planetary atmospheres such as Proxima b. In this study, we presented the results of resolved convection simulations of the Proxima-b environment assuming an Earth-like atmospheric composition, for several incident stellar fluxes and rotation periods. 

For the standard case, organised convection is resolved with a shallow convection of 2.5~km in depth and 60~km in diameter. Convective clouds are present reaching up to 20~km of altitude and 150~km in diameter. Inside this strong updraft, vertical velocity can reach 40~m~s$^{-1}$ with surface rain values of 120 mm/day under. A thick 100~$\%$ cloud cover between 12 and 18~km is present. The convective clouds transport water vapor from the surface to the troposphere. Around 19~km, this enrichment of water vapor is maximal with a four-fold increase compared to initial conditions.

Simulations with higher stellar fluxes and slower rotation rate were conducted. The bond albedo is sensitive to these two physical parameters. The albedo increases when the stellar flux increases, from an increase of the low altitude cloud fraction and a decrease of the depth of the high-altitude cloud layer, overall increasing the albedo.
The decrease of the rotation rate has the effect to change the general circulation especially between 20 and 30~km. This difference of circulation triggers a thicker high-altitude cloud layer thus increasing the bond albedo.

The short wave cloud feedback resolved by the model is lower than the cloud feedback predicted by previous GCM simulations \citep{Yang13,Yang19}. The shallow convection and low altitude clouds are different in the two models, in the CRM the shallow convection is thicker and the cloud coverage of low-altitude clouds is lower leading to a smaller albedo. The high altitude cloud coverage is similar between the two models. While the CRM exhibits a small overestimation 
\bigbreak
of the shallow convection height, this discrepancy between the cloud coverage obtained with this model and the cloud coverage from the LMD GCM is suggesting that the moist convection parametrization scheme does not handle correctly the convection and cloud formation of the near surface area. This study will be used to calibrate a more realistic convection representation such as a thermal plume model \citep{Rio08,Rio10}. 

\bigbreak

The simulations show signs of organized convection with cellular cluster of about 150~km in diameter. However the domain size could be one limit to the aggregation of convection. In such incoming stellar environment, it is important to consider the albedo effect of aggregation of convection .
Simulations with a larger horizontal domain should be performed to fully understand the mechanisms of that phenomenon \citep{Bret05,Wing14}. 

The high velocity inside the largest convective clouds with water cycle is the perfect environment for the occurrence of lightning. NOx and HOx chemistry could therefore be triggered and have an impact on stellar absorption \citep{Arda17} 
\bigbreak
and on HCN production rate, a key feedstock of life's building blocks \citep{Rimm19}. To investigate these effects, improvement is needed into the representation of the water cycle in the microphysical model that would need to handle separately liquid water cloud and ice water cloud as well as the inclusion of graupel. 

The simulations here are for an atmospheric composition close to the Earth. Thanks to the versatility of the model the impact of different composition on the cloud coverage will be explored, for example the impact of CO$_2$ abundance, thought to decrease stratocumulus cloud coverage \citep{Schn19}. Water vapour has been detected on the habitable-zone super-Earth exoplanet K2-18~b in an atmosphere thought to be hydrogen-rich \citep{Benn19,Tsia19}. The difference of molecular mass between hydrogen and water will engender compositional buoyancy that will impact the convection \citep{Leco17,Char21}. Another compositional effect that will be explored is the non-dilute convection regime \citep{Ding16}, where the condensing component is the principal or non negligible species.

\section*{Acknowledgements}

Authors thank the anonymous reviewer that help improve the paper. Authors thank Caroline Muller and Aymeric Spiga for helpful comments and discussions. This project has received funding from the European Research Council (ERC) under the European Union’s Horizon 2020 research and innovation program (grant agreement No. 740963/EXOCONDENSE). This project has received funding from the European Union’s Horizon 2020 research and innovation program under the Marie Sklodowska-Curie Grant Agreement No. 832738/ESCAPE. This work was granted access to the High-Performance Computing (HPC) resources of Centre Informatique National de l’Enseignement Supérieur (CINES) under the allocations n°A0020101167 and A0040110391 made by Grand Équipement National de Calcul Intensif (GENCI). This work was granted access to the HPC resources of the institute for computing and data sciences (ISCD) at Sorbonne Université. This work benefited from the IPSL ciclad-ng facility. M.T. thanks the Gruber Foundation for its generous support to this research. This work has been carried out within the framework of the National Centre of Competence in Research PlanetS supported by the Swiss National Science Foundation. M.T. acknowledges the financial support of the SNSF.

\appendix
\begin{appendices}
\section{TOGA-COARE modeling}
\label{App}

\begin{center}
 \begin{figure}
 \includegraphics[width=17cm]{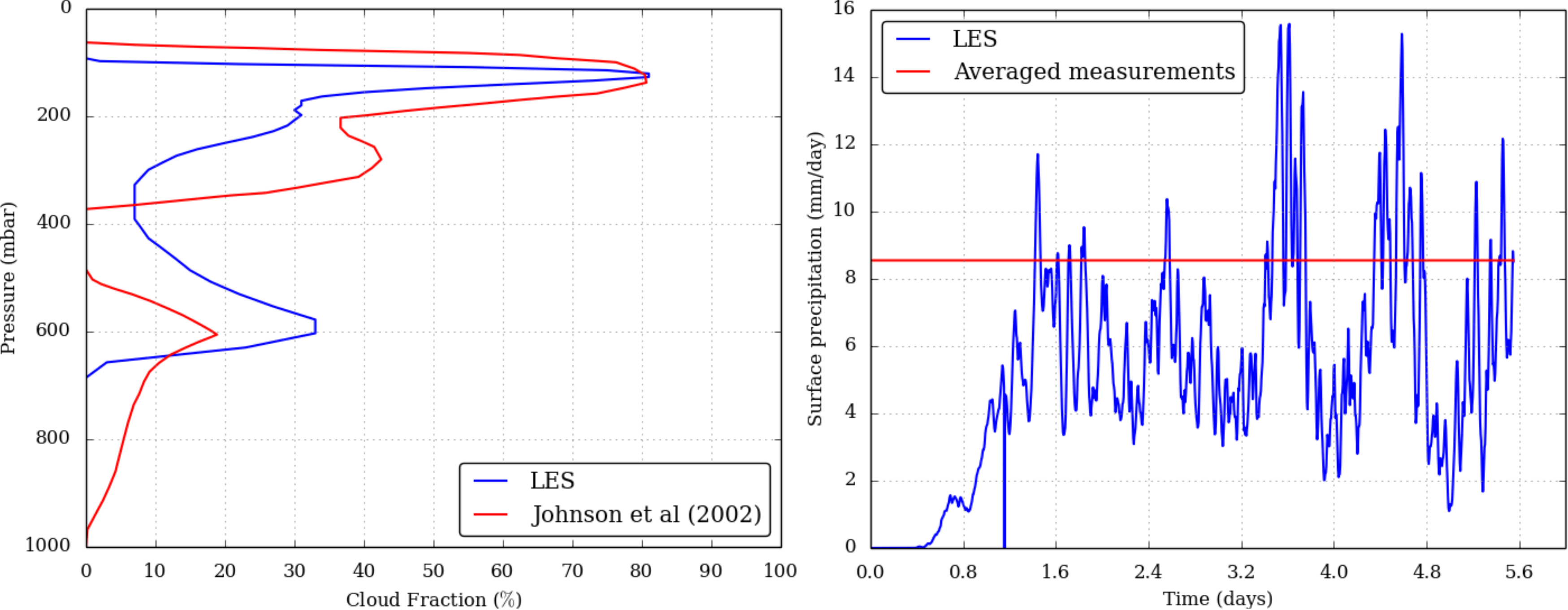}
 \caption{Left : average cloud fraction resolved by the model (CRM) and by \cite{John02}. Right : surface rain(mm/day) over the whole domain (CRM) and averaged measurements. }
 \label{A1}
\end{figure}
\end{center}

As stated in Section~\ref{Sec:model} the model presented in this study was tested with Earth tropics convection data from the TOGA-COARE campaign. The temperature, pressure, water vapor and wind profiles are taken from the measurement campaign. Advection heating and water vapour advection are also taken account. The surface temperature is set up at the beginning of the simulations and then is free to evolve during the simulation. With similar settings that those described in Section~\ref{Sec:model} with a Sun spectrum and a diurnal cycle, the model resolves convection showed at Fig~\ref{A1} after 6 Earth days of simulation.

The cloud fraction (left) is calculated using \cite{Xu1991} formulations to be compared to \cite{John02}. The overall behaviour is very similar, however the low altitude clouds are slightly overestimated by the model and the high altitude clouds are a little underestimated. The height of the cloud bottom boundary, about 750~hPa, is as well overestimated compared to similar modeling, about 950~hPa \citep{Dale15}. The surface rain resolved by the model is close to the average of models from \cite{John02} and \cite{Wang13}. However the model is not designed to reproduce extreme events, such as strong sudden increase of a domain average surface rain over 40 mm/day \citep{Wang13}. This model is suited to exhibits realistic behaviour for Earth tropics convection.

\end{appendices}

\newpage

\bibliographystyle{apalike}

\end{document}